\documentclass[apjl]{emulateapj}

\shorttitle{IR spectra of HBC$^+$ and HBC$^{2+}$}
\shortauthors{Zhen et al.}

\begin{document}
\title{Infrared Spectra of Hexa-peri-hexabenzocoronene cations: HBC$^+$ and HBC$^{2+}$}

\author{Junfeng Zhen$^{1,2,*}$, Pablo Castellanos$^{1,2}$, Jordy Bouwman$^{2,3}$, Harold Linnartz$^{2}$, Alexander G. G. M. Tielens$^{1}$} 

\affil{$^{1}$Leiden Observatory, Leiden University, P.O.\ Box 9513, 2300 RA Leiden, The Netherlands}   
\affil{$^{2}$Sackler Laboratory for Astrophysics, Leiden Observatory, Leiden University, P.O.\ Box 9513, 2300 RA Leiden, The Netherlands}
\affil{$^{3}$IMM, FELIX Lab, Radboud University Nijmegen, Toernooiveld 7c, 6525 ED Nijmegen, The Netherlands}
\email{zhen@strw.leidenuniv.nl}

\begin{abstract}
We present the first infrared (IR) gas phase spectrum of a large and astronomically relevant PAH cation (C$_{42}$H$_{18}$$^+$, HBC$^+$) and its di-cation (C$_{42}$H$_{18}$$^{2+}$, HBC$^{2+}$). The spectra are recorded via infrared multi-photon dissociation (IRMPD) spectroscopy of ions stored in a quadrupole ion trap, using the intense infrared radiation of a free electron laser in the 530$-$1800 cm$^{-1}$ (5.6$-$18.9 $\mu$m) range. HBC$^{+}$ shows main intense absorption peaks at 762 (13.12), 1060 (9.43), 1192 (8.39), 1280 (7.81), 1379 (7.25) and 1530 (6.54) cm$^{-1}$($\mu$m), in good agreement with DFT calculations that after scaling to take the anharmonicities effect into account. HBC$^{2+}$ has its main absorption peaks at 660 (15.15), 766 (13.05), 1054 (9.49), 1176 (8.50), 1290 (7.75), 1370 (7.30) and 1530 (6.54) cm$^{-1}$($\mu$m). Given the similarity in the cationic and di-cationic spectra, we have not identified an obvious diagnostic signature to the presence of multiply charged PAHs in space. While experimental issues associated with the IRMPD technique preclude a detailed comparison with interstellar spectra, we do note that the strong bands of HBC$^+$ and HBC$^{2+}$ at $\sim$ 6.5, 7.7, 8.4 and 13.1 $\mu$m coincide with prominent aromatic infrared bands (AIBs). HBC has only trio CH groups and the out-of-plane CH bending mode of both HBC cations is measured at 13.1 $\mu$m, squarely in the range predicted by theory and previously found in studies of small (substituted) PAHs. This study therefore supports the use of AIBs observed in the 11$-$14 $\mu$m range as a diagnostic tool for the edge topology of large PAHs in space.

\end{abstract}


\section{Introduction}
\label{sec:intro}
The aromatic infrared bands (AIBs) dominate the near and mid IR spectra of many interstellar sources. These bands are generally attributed to IR fluorescence of large ($\sim$ 50 C atoms) polycyclic aromatic hydrocarbon (PAH) molecules after excitation by ultraviolet (UV) photons \citep{all89,pug89,sel84,gen98,tie13}. PAHs are expected to form in the ejecta of AGN stars and are ubiquitous and abundant, containing $\sim$ 10\% of the elemental carbon in the interstellar medium (ISM). They play an important role in the ionization and energy balance of the ISM \citep[and references therein]{tie08}. The assignment of the AIBs to PAH molecules has been discussed in several studies \citep{all89,peeters2002,tie05,tie08}. Generally, the prominent 3.3 $\mu$m band is linked to aromatic CH stretches, the 6.2 and 7.7 $\mu$m features are considered to be due to CC vibrations of aromatic rings, and the 8.6 $\mu$m as well as the longer wavelength region bands are assigned to CH in- and out--of--plane bending and/or CCC bending vibrations \citep{tie13}. 

IR spectroscopy provides a powerful tool to study the molecular characteristics of the emitting species. Specifically, for the CH out$-$of$-$plane (oop) bending vibrations in the 11$-$14 $\mu$m region, a different number of adjacent CH bonds gives rise to vibrations in distinctly different wavelength regions \citep{hon01,tie08}. These spectroscopic rules on H-adjacency and the positions of the oop modes have largely been derived from experimental studies on small (substituted) aromatic systems such as benzene and naphthalene \citep{bellamy} and are supported by Density Functional Theory (DFT) calculations \citep[c.f.][]{bauschlicher2010}. Furthermore, laboratory and quantum chemical studies have revealed that the CC stretching modes in the 6$-$9 $\mu$m region are much stronger in ionized PAHs than in their neutral counterparts \citep{allamandola1999,peeters2002}. These three bands (6.2, 7.7, and 8.6 $\mu$m), which correlate well, therefore {\bf may be attributed} to PAH cations. Recent studies, using PAH ions trapped in hydrogen matrices {\citep{tsu16}} or  measured in tagged ion dissociation studies \citep{rick09,har09} also show nice agreement with vibrational modes recorded for protonated PAH species. This is supported by observational studies that reveal that these bands grow in strength -- relative to the CH bands characteristic for neutral PAHs -- in regions characterized by high ionization parameters \footnote{the ionization parameter is given by the ratio of the ionization rate over the recombination rate and scales with $G_0T^{1/2}/n_e$ with $G_0$ the strength of the radiation field in terms of the average interstellar radiation field, $T$ the temperature, and $n_e$ the electron density \citep{tie05}.}; e.g., close to the illuminating source \citep{berne2007,rapacioli2005,boersma2013}. However, in such regions, PAH cations can also be further ionized, resulting in doubly, triply, or even higher charged ions. In competition with the ionization channel, PAH cations can also fragment upon energetic photo-excitation. This fragmentation follows the available pathways (i.e., dehydrogenation or loss of C$_{2}$ or C$_{2}$H$_{2}$ units), producing PAH derivative species along a top-down chemical scenario \citep{zhen2014, bou16, pet16}. Photo-ionization and photo-fragmentation compete, and their relative importance depends both on the geometry of PAH molecules and the excitation wavelengths used \citep{zhen2015, zhe2016}. 

In order to elucidate possible AIB carriers, neutral and ionized PAH molecules with different size and molecular geometries have been studied, both in the laboratory and by theory. Experimentally, IR spectra of PAH neutrals, cations and other derivatives have been measured in rare gas matrices \citep[e.g.][]{szc931, szc932, hud951, hud952, mat03, ber07, tsu16}. In the gas phase IR emission spectra have been studied for neutral PAHs and PAH cations \citep{coo98, kim01, kim02}. Absorption gas phase spectra of PAH cations have been studied using IRMPD \citep{oom01, oom06} or messenger atom photo-dissociation (MAP) spectroscopy \citep{pie99, dun03, dou08, rick09}. A recent application shows that this method is even suited to study fullerene cations \citep{cam15,kuh16}. While some PAH spectral bands coincide approximately with prominent AIBs, the agreement is not always compelling. A more convincing overlap with astrophysical emission spectra is obtained with composite spectra that consist of (selected) distributions of spectra of different PAH and PAH$^{+}$ species that have been recorded using matrix isolation spectroscopy or computed using DFT calculations \citep{peeters2002, cami2011}.

For technical reasons, IR laboratory studies of PAHs have largely been focused on smaller species, containing up to 24 C-atoms. IR spectra of large interstellar relevant PAHs ($\sim$ 50 C-atoms) are generally lacking and this is also definitely the case for their cations or di-cations. In this article, we present the first IR spectra of the cation and di-cation of hexa-peri-hexabenzocoronene: C$_{42}$H$_{18}^+$ and C$_{42}$H$_{18}$$^{2+}$. The resulting IR spectra are compared with quantum chemical computations and astronomical data. HBC (Figure 1) has been selected as it is an all-benzenoid PAH with a size in the astrophysically-relevant range \citep{tie08,croiset2016} that possibly serves as a prototypical example for large(r) PAHs. Moreover, the molecular geometry of this PAH is special, as it has only CH groups with a \textquotedblleft triply-adjacent\textquotedblright or \textquotedblleft trio\textquotedblright CH-structure, as shown in Figure 1. This allows us to unambiguously identify the wavelength of the out-of-plane CH bending mode(s) for such a geometry. 

\section{Experimental Methods}
\label{sec:exp}
The IR spectra of the fully-benzenoid hexa-peri-hexabenzocoronene cation (C$_{42}$H$_{18}^{\phantom{26}+}$ with $m/z = 522.14$) and its di-cation (C$_{42}$H$_{18}$$^{2+}$) with $m/z = 261.07$) have been recorded using our instrument for Photo-dissociation of PAHs (i-PoP) connected to the Free Electron Laser for InfraRed eXperiments (FELIX) at Radboud University \citep{oep95}. The apparatus comprises a quadrupole ion-trap (QIT) time-of-flight (TOF) mass spectrometer and has been described in detail in \citet{zhe14}. Below only a brief description of the relevant details is given.

Neutral HBC precursor molecules are transferred into the gas phase by heating HBC powder (Kentax, purity higher than 99.5 \%) in an oven ( $\sim$ 580K) until it slowly evaporates. 
The sublimated HBC molecules are ionized using electron impact ionization (83~eV electron impact energy). The resulting HBC$^+$ and HBC$^{2+}$ cations are transported into the ion trap via an ion gate and are trapped by applying a 1~MHz radiofrequency electric field of 3150 and 2130 V$_{p.p.}$, respectively, onto the ring electrode. Helium is introduced continuously into the trap to thermalize the ion cloud through collisions.

Spectral information is obtained over the 530 to 1800~cm$^{-1}$ range by IRMPD using FELIX. This free electron laser delivers 5~$\mu$s long macropulses of light at a repetition rate of 10~Hz with a macropulse energy up to 100~mJ. The bandwidth of the laser is Fourier limited and is typically 0.6\% (FWHM) of the central wavelength. When the laser wavelength is in resonance with an allowed vibrational transition in the ion, the absorption of multiple photons takes place, causing the molecule to dissociate. The resulting fragments are then pulse-extracted from the ion trap and analyzed in a TOF mass spectrometer. All channels corresponding to photo-fragmentation of the parent are summed up and normalized to the total signal, i.e. parent plus photo-fragment ions in the trap, yielding the relative photo-fragmentation intensity. The infrared spectrum of the ion is constructed by plotting the fragment ion yield as a function of wavelength. The recorded spectra are normalized to the {\bf pulse energy} to correct for any power dependencies. 

Mass spectra data for HBC$^+$ and HBC$^{2+}$ are recorded typically with steps of 5 cm$^{-1}$. Optimal irradiation times amount to 0.8 s (typically $\sim$ 8 pulses). In order to obtain a good signal-to-noise ratio, mass-spectra are averaged 25 times. In addition, blank spectra are recorded in order to subtract fragmentation peaks that are already present in the background signal. 

\begin{figure}[t]
  \centering
  \includegraphics[width=\columnwidth]{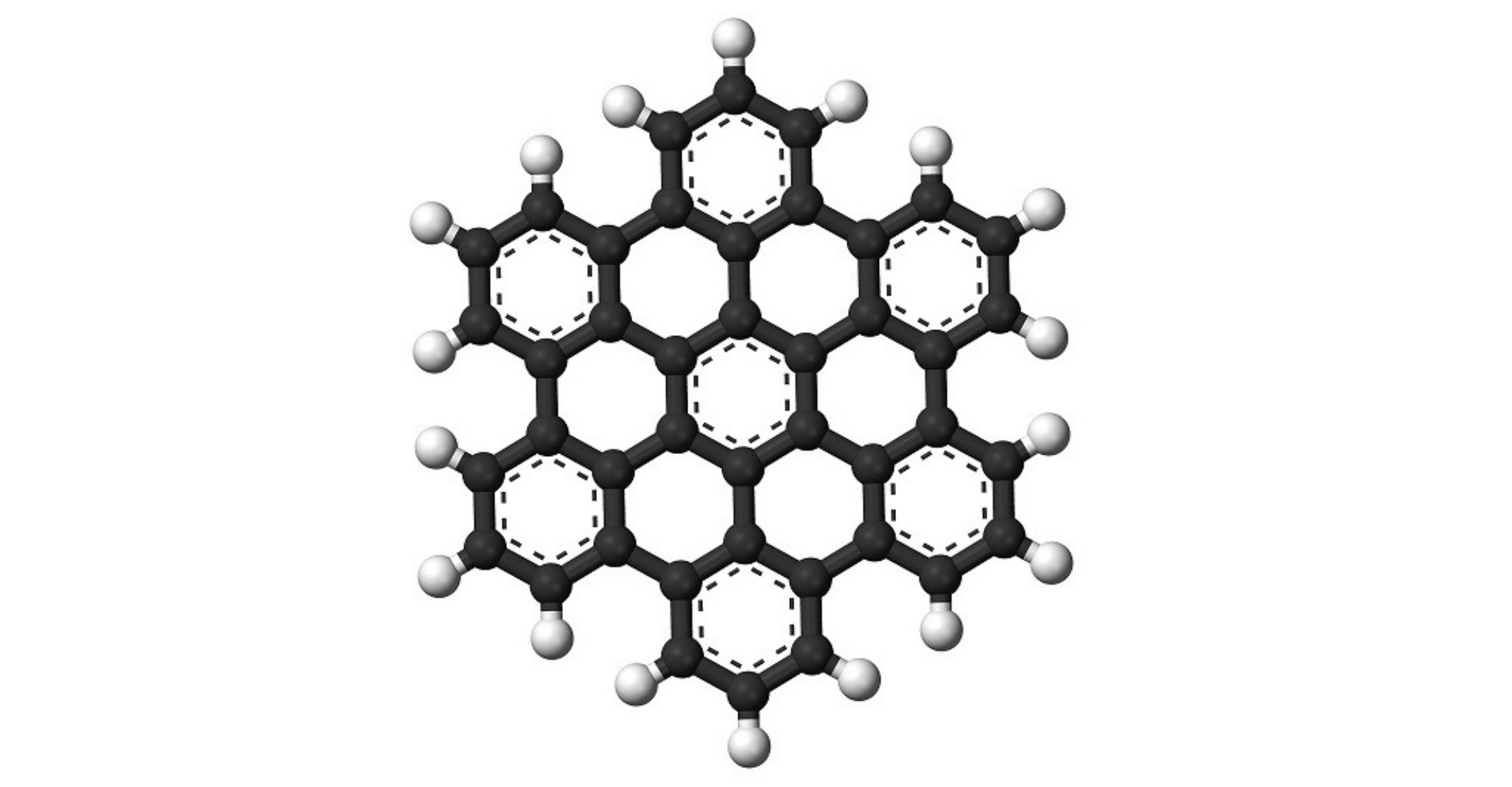}
  \caption{The molecular geometry of hexa-peri-hexabenzocoronene (HBC, C$_{42}$H$_{18}$).
  }
  \label{fig1}
\end{figure}

\begin{figure*}[t]
  \centering
  \includegraphics[width=\textwidth]{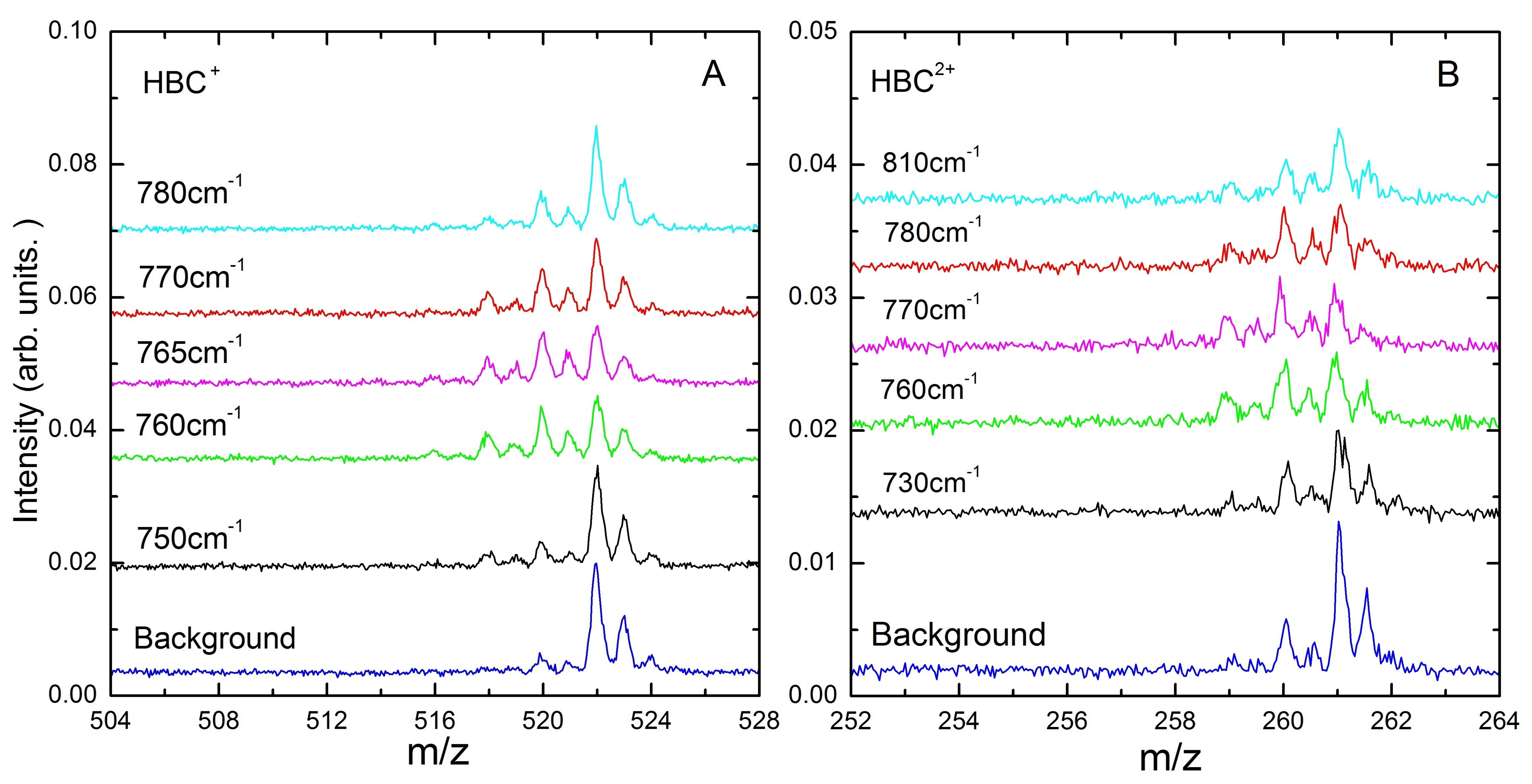}
  \caption{Mass spectrum of the photo-products resulting from irradiation of HBC$^+$ and HBC$^{2+}$: (A) The photo-dehydrogenation behavior of HBC$^+$, irradiated at different wavelength in the 762cm$^{-1}$ band range; (B) The photo-dehydrogenation behavior of HBC$^{2+}$, irradiated at different wavelengths in the 766cm$^{-1}$ band range. Additional peaks in the background trace are isotopes and fragmentations induced by the electron gun.
  }
  \label{fig2}
\end{figure*}

\begin{figure*}[t]
  \centering
  \includegraphics[width=\textwidth]{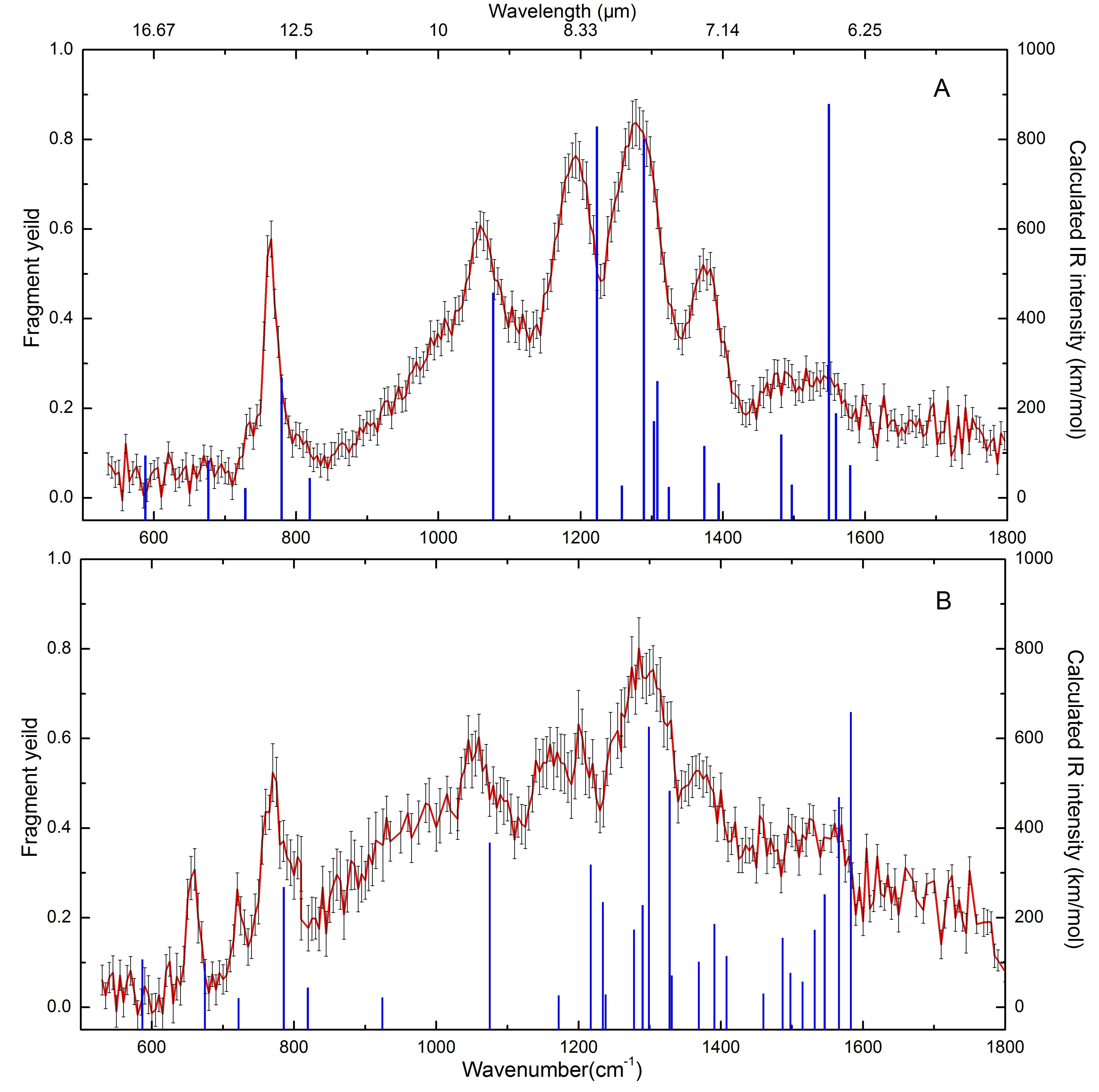}
  \caption{Measured IRMPD spectrum of gas-phase HBC$^+$ and HBC$^{2+}$. Computed vibrational normal modes scaled to correct for anharmonicities (taken from \citet{boe2014}) are represented by vertical bars.
  }
  \label{fig3}
\end{figure*}

\begin{figure}[t]
  \centering
  \includegraphics[width=\columnwidth]{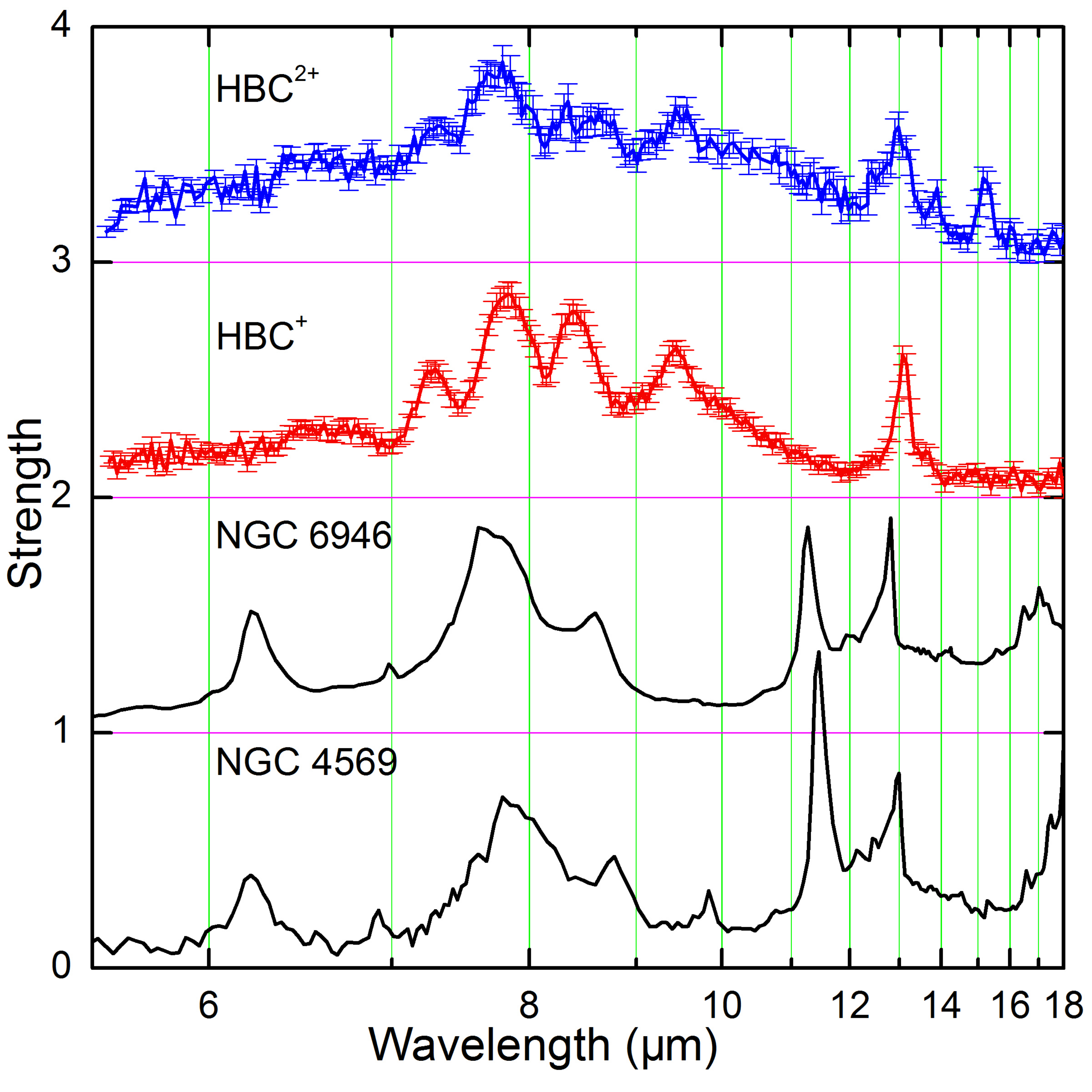}
  \caption{IR spectra of HBC$^+$ and HBC$^{2+}$ (top) compared to two representative AIB emission spectra (bottom) taken towards NGC 6946 and NGC 4569 and adopted from \citet{smi07}.
  }
  \label{fig4}
\end{figure}

\section{Results and discussion}
\label{sec:results}
Figure 2 shows the mass spectra of the photo-products that result upon excitation of the 762~cm$^{-1}$ band of HBC$^+$ (panel A) and the 766 cm$^{-1}$ band of HBC$^{2+}$ (panel B). The mass spectra before irradiation (labeled as background) reveal a small amount of residual fragmentation due to the electron impact ionization and also include $^{13}$C isotopes that contribute to the main mass peak signal. The different graphs in Fig. 2(A) show the fragmentation patterns of trapped HBC$^+$ for a set of wavelengths between 750 and 780~cm$^{-1}$. Several dehydrogenated fragment ions can be seen that are the consequence of multiple and cumulative photon absorptions. The photo-dissociation pattern follows mainly sequential 2H (or H$_2$) loss channels, similarly to what was observed for HBC$^+$ in \citet{zhe14} upon irradiation with optical photons (532nm). Figure 2(B) shows the same data for HBC$^{2+}$, now after irradiation with wavelengths between 730 and 810 cm$^{-1}$. Similar to the photo-fragmentation behavior of HBC$^+$, a wide range of partially dehydrogenated fragment ions is found and also in the case of HBC$^{2+}$ the photo-dissociation pattern mainly follows sequential 2H (or H$_2$) losses.

Figure 3 shows the resulting IR spectra for HBC$^+$ (panel A) and HBC$^{2+}$ (panel B) from 530 to 1800 cm$^{-1}$ (5.6$-$18.9 $\mu$m). The spectra are recorded by integrating the fragmentation signals that vary with wavelength as illustrated in Figure 2. This figure shows a number of clearly resolved vibrational bands, with reasonable signal-to-noise ratios, even in the case of HBC$^{2+}$. Many of the bands are rather broad, which is commonly observed for IRMPD spectra and is attributed to anharmonic effects induced by the multiphoton process. It is also possible that broad unresolved bands hide underneath the separate bands visible here. This broadening may be less prominent in experiments using rare gas ion tagging. Such experiments, however, may come with the drawback that transitions substantially shift upon complexation, depending on where the atom tags to the PAH. So far, the existing tagging experiments are available mainly for smaller PAH-cations, and likely also hard to apply to di-cations. An extension to larger species will be needed to directly compare with the results presented here.
 
Also included in Figure 3 are the theoretical predictions of the vibrational band positions of HBC$^+$ and HBC$^{2+}$ \citep{boe2014} that have been scaled by a uniform factor of 0.958 to correct for anhamonicities\footnote{This global scaling factor corrects approximately for the shifts in peak position due to anharmonicity at 0K (Langhoff 1996). This is an approximation as individual bands show different anharmonic corrections in the calculations (Mackie et al 2016). We also stress that this does not account for the anharmonic shifts in peak position of highly excited species inherent to the IRMPD technique (Oomens et al 2003).}. In the case of HBC$^{2+}$ we considered both the singlet and triplet states, yet the singlet state is chosen for comparison as it is lowest in energy \citep{boe2014}. Table~1 summarizes all vibrational and computed wavelengths with their assignments. The vibrational peak positions are determined directly from the maximum intensity without fitting the overall spectrum. The latter prohibits an overfitting of our data as we do not know how many broad bands may contribute to the overall profile. 

Fig. 3(A) and Table 1 compare our experimental IRMPD HBC$^+$ spectrum with the theoretical predictions \citep{boe2014}. The most prominent bands include the carbon ring distortions at 1280 (7.81), 1379 (7.25) and 1530 (6.54) cm$^{-1}$ ($\mu$m), and the CH in- or out-of-bending modes at 762 (13.12), 1060 (9.43) and 1192 (8.39) cm$^{-1}$ ($\mu$m). The intense band at 762 cm$^{-1}$ with a FWHM of $\sim$ 20 cm$^{-1}$ in the IRMPD spectrum corresponds with the 779 cm$^{-1}$ feature in the DFT spectrum, which arises from out-of-plane CH bending modes. The bands measured at 1060 and 1192 cm$^{-1}$ are close to transitions predicted at 1077 and 1223 cm$^{-1}$, which arise from in-plane CH bending modes. The experimental bands at 1280 and 1379 cm$^{-1}$ are close to the theoretically predicted bands at 1289 and 1374 cm$^{-1}$. It is possible that the 1280 cm$^{-1}$ band also includes bands at 1303 and 1308 cm$^{-1}$. All these bands have a FWHM of $\sim$ 60 cm$^{-1}$. The band at 1530 cm$^{-1}$ is broad -- with a FWHM of $\sim$ 120 cm$^{-1}$ -- and likely is composed of a number of bands predicted at 1482, 1549, 1559 and 1579 cm$^{-1}$ (Table 1).

Figure 3(B) and Table 1 provide similar information for HBC$^{2+}$. The error bars here are larger due to a lower ion signal, as fewer di-cations are trapped in the QIT. The most prominent bands include again the carbon ring distortions at 1290 (7.75), 1370 (7.30) and 1530 (6.54) cm$^{-1}$ ($\mu$m), and the CH in- or out-of-bending modes at 660 (15.15), 766 (13.05), 1054 (9.49) and 1176 (8.50) cm$^{-1}$ ($\mu$m). The bands at 660 and 766 cm$^{-1}$ with FWHMs of $\sim$ 20 cm$^{-1}$ in the IRMPD spectrum correspond with the bands at 675 and 785 cm$^{-1}$ in the DFT spectrum, {\bf which arise} from out-of-plane CH bending modes. The band measured at 1054 cm$^{-1}$ is close to the transition predicted at 1075 cm$^{-1}$, and the band at 1176 cm$^{-1}$ in the IRMPD spectrum links to bands at 1217 and 1234 cm$^{-1}$, which arise from in-plane CH bending modes. The band measured at 1290 cm$^{-1}$ corresponds to the bands computed at 1299 and 1328 cm$^{-1}$ and possibly also contains contributions from nearby bands at 1278, 1290 and 1328 cm$^{-1}$. The band at 1370 cm$^{-1}$ in the IRMPD spectrum is found in the DFT calculations at 1391 cm$^{-1}$ and possibly also contains contributions from the computed 1369 and 1408 cm$^{-1}$ bands (Table 1). All these bands have a FWHM of $\sim$ 60 cm$^{-1}$. The {\bf band at 1530 cm$^{-1}$ is broad band} with a FWHM of $\sim$120 cm$^{-1}$ and likely contains contributions from other bands calculated at 1487, 1532, 1546 1566 and 1583 cm$^{-1}$ (Table 1). 

As a general point, we emphasize that, despite the fact that a scaling factor of 0.958 is already applied, all experimentally measured band positions of HBC$^+$ and HBC$^{2+}$ are red-shifted with respect to the calculated positions. We attribute this shift to the large anharmonicity associated with the high internal excitation required for the dissociation of large PAHs, which is inherent to the IRMPD technique \citep{oom03}. We note that these shifts are different for different vibrational modes. For example, from Fig. 3(A) and Table 1, it becomes clear that the 1192 cm$^{-1}$ band is about $\sim$ 30 cm$^{-1}$ situated to lower energy than predicted, whereas for the 1280 cm$^{-1}$ band the shift is only $\sim$10  cm$^{-1}$. Taking this into account, the agreement between the experimental values presented here and the available theoretical data is reasonable. 

While overall, the comparison between experiments and theory is very encouraging, there are also some issues. The HBC$^+$ and HBC$^{2+}$ bands at 1060 and 1054 cm$^{-1}$, respectively, exhibit seemingly broad wings towards lower frequencies in the 850$-$1000 cm$^{-1}$ range. Theory does not predict vibrational normal modes in this region, and these broad bands may be examples of the extreme effects of anharmonicity on IRMPD spectra. As another discrepancy with theory, we note that theory predicts bands at 588 and 587 cm$^{-1}$ for HBC$^+$ and HBC$^{2+}$, respectively, which are not apparent in the experiments. Perhaps, anharmonicity is also a factor here, shifting these bands outside of the spectral range scanned with FELIX. In addition, the non-linear intensity response of IRMPD in combination with a lower output power at the low (and high side) of the accessible wavelength domain may be also responsible for this observation.

One further difference occurs at high frequencies where DFT calculations predict a strong band for HBC$^+$ at 1549 cm$^{-1}$ and for HBC$^{2+}$ at 1583 cm$^{-1}$  with a number of relatively weaker bands nearby (Table 1). In contrast, rather than one strong band, the experiments reveal weak, but broad features between 1430 and 1630 cm$^{-1}$. This difference may again reflect extreme anharmonicity for these modes. Likewise, we do detect the theoretically predicted band at 660 cm$^{-1}$ of HBC$^{2+}$ but a similarly predicted band of HBC$^+$ is missing in our spectrum. On the one hand, this may indicate an issue with the theoretical spectrum of HBC$^+$. In addition, this might result from a much larger anharmonicity in HBC$^+$ (than in HBC$^{2+}$) which would shift this band rapidly out of resonance in our experiment before fragmentation can occur. Further experiments and quantum chemical calculations are required to improve our understanding of these observations.

\begin{table*}
\centering
\caption{Infrared Transitions Measured and Calculated for HBC$^+$ and HBC$^{2+}$
\label{BenchCalcs}}
\begin{tabular}{llllllllll}
\hline
\multicolumn{3}{c}{HBC$^+$}&\multicolumn{3}{c}{HBC$^{2+}$}&\\

Experimental & Computed & Computed & Experimental & Computed & Computed & Assignment\\
band&band&band&band&band&band&&\\
cm$^{-1}$, $\mu$m & cm$^{-1}$& km$/$mol& cm$^{-1}$, $\mu$m & cm$^{-1}$& km$/$mol &\\
\hline
& 262& 33&&264& 23&\\
& 588& 94&&587& 106&\\
& 676& 82$^c$& 660 (m, A), 15.15& 675& 98&out-of-plane CH bending\\
732, (w$^a$, A$^b$), 13.66& 728& 21&720 (w, B), 13.89& 722& 20&out-of-plane CH bending\\
762 (s, B), 13.12& 779 & 267&766 (s, A), 13.05& 785& 268&out-of-plane CH bending\\
&819& 43&804 (w, B), 12.44& 819& 43&out-of-plane CH bending\\
&&&&924& 21&\\
1060 (s, B), 9.43& 1077& 457&1054 (s, B), 9.49& 1075& 367&in-plane CH bending\\
&&&&1172&26&\\
1192 (s, B), 8.39& 1223& 828&1176 (s, B), 8.50& 1217& 318&in-plane CH bending\\
&&&&1234& 234&\\
&&&&1238& 28&\\
&1258& 26&\\
&&&&1278& 172&\\
&&&&1290& 227&\\
1280 (s, B), 7.81& 1289& 800&1290 (s, B), 7.75& 1299& 625& carbon ring distortions\\
&1303& 170&\\
&1308& 259&\\
&1324& 23&&1328& 482&\\
&&&&1331& 70&\\
&&&&1369& 101&\\
1379 (s, B), 7.25& 1374& 115&1370 (s, B), 7.30&  1391& 185& carbon ring distortions\\
&1394& 32&\\
&&&&1408& 114&\\
&&&&1460& 30&\\
&1482& 140&&1487& 154&\\
&1497& 29&&1498& 76&\\
&&&&1515& 57&\\
&&&&1532& 172&\\
&&&&1546& 251&\\
1530 (m, C), 6.54& 1549& 878&1530 (m, C), 6.54& 1566& 468&carbon ring distortions\\
&1559& 188&&1583& 658&\\
&1579& 72&\\
\hline
\end{tabular}
\\
{$^a$}Relative intensities of experimental bands are indicated as w, m, s for weak, medium, strong, respectively;
{$^b$}The uncertainty in the experimental band positions is estimated - following the NIST definition - to be A (1$\sim$3 cm$^{-1}$), B (3$\sim$6 cm$^{-1}$) and C (15$\sim$30 cm$^{-1}$);
$^c$ Vibrational frequencies were scaled by a factor of 0.958 to allow comparison with the experimental data, consistent with the recommended scaling for this level of theory \citep{sco96}, and intensities (km$/$mol) are given in parentheses. We only list bands with intensities larger than 20 km$/$mol.
\end{table*}

\section{Astrophysical relevance and conclusions}
\label{sec:astro}
This paper presents the first IR spectrum of a PAH cation with an astrophysically relevant size -- C$_{42}$H$_{18}$ -- in the gas phase. In addition, this is the first time that the spectrum of a PAH di-cation has been measured experimentally in the gas phase. The experimental IR spectra for both HBC$^+$ and HBC$^{2+}$ (top two traces) are shown together with typical AIB emission spectra (bottom two traces) in Fig.~4. The astronomical emission spectra are from NGC 6946 and NGC 4569 \citep{smi07}. This interstellar spectrum shows the well-known prominent broad AIBs at 6.2, 7.7, 8.6, 11.2 and 12.7 $\mu$m. As shown in the figure, the experimental spectra of HBC$^+$ and HBC$^{2+}$ exhibit similarities to several of the key bands in the AIB spectrum. The strong bands at 7.7 and 8.6 $\mu$m in the laboratory spectrum closely match the positions of the AIB bands at these wavelengths, and the 13.1 $\mu$m band is very close to the 12.7 $\mu$m band due to $"$trio$"$ CH-structure bending modes in HBC. The 11.2 $\mu$m band that shows up in the astronomical spectrum is absent in the HBC data, as this band is due to ``solo'' CH-structure excitation \citep{hon01} and this group is not present in HBC. As such, this comparison also illustrates that the astronomical spectra may be a combination of emission features originating from many separate species, and likely HBC$^+$ and HBC$^{2+}$ will contribute but not at a level that allows an unambiguous identification.

Theoretical and experimental studies have highlighted the large spectral differences between neutral and ionized PAHs (Langhoff 1996; Allamandola et al 1999). As charging involves the delocalized $\pi$ electron system (rather than the localized $\sigma$ electrons), bond strength and mode distribution do not differ much between these charge states. The change in charge distribution does influence, though, the change in dipole moment during the vibration and this is the underlying cause for the striking spectral differences. Our experiments show that -- in contrast to the neutral-cation behavior -- there is no clear spectral signature that can lead to the unambiguous identification of di-cations in space.  This is unfortunate, as a tool for determining the charge state of PAHs in space would contribute substantially to our knowledge on how PAHs behave upon irradiation. The main difference occurs in the 600-700 cm$^{-1}$ region where a band at 660 cm$^{-1}$ in HBC$^{2+}$ is present in both our experimental study and in the DFT calculations but the corresponding theoretical band in HBC$^+$ is not observed in the IRMPD study. As argued above, this may reflect a difference in anharmonicity between HBC$^{2+}$ and HBC$^+$ -- rather than a spectral signature of di-cations -- which adversely affects the measured spectra. Further studies will be required to address this issue.

The strong transitions in the 530$-$1800 cm$^{-1}$ (5.6$-$18.9 $\mu$m) IR fingerprint region have been compared to published DFT data and we conclude that, despite drawbacks that complicate the interpretation of band profiles and band intensities, the applied IRMPD technique provides useful band positions for IR active modes. Specifically, we emphasize that the measured HBC spectra support the assignment \citep{hon01} of the interstellar 12.7 $\mu$m band to the CH out-of-plane bending mode associated with trio's on (large) PAHs (although a duo assignment for the interstellar band may also be viable; \citep{bauschlicher2010}). Finally, we stress that this study demonstrates the potential of IRMPD spectroscopy  for much larger species than measured before and this work paves the way to systematically study even larger PAHs ($>$60 C-atoms) with our set-up in the future.

\acknowledgments
The authors gratefully acknowledge the FELIX staff for their technical assistance during preparation and the actual beamline shifts. We are grateful to M.J.A.\ Witlox for technical support. We are grateful to Giacomo Mulas and Tao Chen for many very useful discussions. Studies of interstellar chemistry at Leiden Observatory are supported through advanced-ERC grant 246976 from the European Research Council, through a grant by the Netherlands Organisation for Scientific Research (NWO) as part of the Dutch Astrochemistry Network, and through the Spinoza premie. JB acknowledges NWO for a Veni grant (722.013.014). We acknowledge the European Union (EU) and Horizon 2020 funding awarded under the Marie Sklodowska-Curie action to the EUROPAH consortium, grant number 722346.

\end{document}